\documentclass[11pt,twoside]{article}

\usepackage{asp2014}

\aspSuppressVolSlug
\resetcounters

\begin{document}

\title{PSF quality metrics in the problem of revealing Intermediate-Mass Black Holes using MICADO@ELT}

\author{Mariia Demianenko,$^{1,2}$ J\"{o}rg-Uwe Pott,$^1$ and Kai Polsterer$^2$}
\affil{$^1$Max Planck Institute for Astronomy, Heidelberg, Germany; \email{demianenko@mpia.de}}
\affil{$^2$Heidelberg Institute for Theoretical Studies, Heidelberg, Germany}

\paperauthor{Mariia Demianenko}{demianenko@mpia.de}{ORCID_Or_Blank}{Max Planck Institute for Astronomy}{Galaxies and Cosmology Department (GC)}{Heidelberg}{}{69126}{Germany}
\paperauthor{J\"{o}rg-Uwe Pott}{jpott@mpia.de}{}{Max Planck Institute for Astronomy}{Galaxies and Cosmology Department (GC)}{Heidelberg}{}{}{Germany}
\paperauthor{Kai Polsterer}{Author3Email@email.edu}{}{Heidelberg Institute for Theoretical Studies}{Astroinformatics (AIN)}{Heidelberg}{}{}{Germany}



\begin{abstract}
Nowadays, astronomers perform point spread function (PSF) fitting for most types of observational data. Interpolation of the PSF is often an intermediate step in such algorithms. In the case of the Multi-AO Imaging Camera for Deep Observations (MICADO) at the Extremely Large Telescope (ELT), PSF interpolation will play a crucial role in high-precision astrometry for stellar clusters and confirmation of the Intermediate-Mass Black Holes (IMBHs) presence. Significant PSF variations across the field of view invalidate the approach of deconvolution with a mean PSF or on-axis
PSF. The ignoring of PSF variations can be especially unsatisfactory in the case of Single Conjugate Adaptive Optics (SCAO) observations, as these sophisticated and expensive systems are designed to achieve high resolution with ground-based telescopes by correcting for atmospheric turbulence in the direction of one reference star. In plenty of tasks, you face the question: How can I establish the quality of PSF fitting or interpolation? Our study aims to demonstrate the variety of PSF quality metrics, including the problem of revealing IMBHs in stellar clusters. 
\end{abstract}



\section{Intermediate-Mass Black Holes: astrophysical context}

\subsection{Why are they so intriguing?}
The seeding mechanisms and evolution of supermassive black holes (SMBHs) in galaxy centers still remain poorly understood: they were listed among the key astrophysical questions in the Astro-2020 Decadal Survey by the US National Academies of Science. SMBHs grow from yet unidentified BH seeds by accreting gas in the active galactic nucleus (AGN) \citet{2012Sci...337..544V} phase or via coalescences during galaxy mergers \citet{2005LRR.....8....8M}. SMBH binaries form at the pre-coalescence stage and then become the sources of gravitation waves, which will likely be detected at cosmological distances by the LISA space mission \citet{2022arXiv220111633J}. Which growth channel prevails is still a matter of debate. The discovery of quasars in the early Universe ($z > 6.3$) hosting SMBHs as heavy as $10^{10} M_{\odot}$ \citet{mortlock11,2015Natur.518..512W} cannot be explained by gas accretion onto stellar-mass black hole seeds ($\lesssim 100 M_{\odot}$) alone; massive seeds ($>10^5 M_{\odot}$) formed by direct collapse \citet{2004Natur.428..724P} of massive gas clouds provide a viable solution but also raise more questions. Low-mass BH seeds should leave behind a population of intermediate-mass black holes (IMBHs; $100<M_{BH}<10^5 M_{\odot}$) holding the key for the question of SMBH origin \citet{2017IJMPD..2630021M}. 
\subsection{MICADO@ELT: source of future conclusive IMBH sample}
Modern astrophysics provides two conclusive black hole confirmation methods: gravitational lensing (M87 \citet{2019ApJ...875L...1E}, Sgr $A^{*}$ in Milky Way \citet{2022ApJ...930L..12E}) and measuring of stellar velocities around the black hole (\citet{2008ApJ...689.1044G,2010RvMP...82.3121G}). MICADO@ELT aimed at the second method and having an unprecedented resolution of 4 mas/pixel at near-IR bands in one of the regimes. Using high-resolution multi-epoch images astronomers will be able to measure proper motions (tangential velocities), resolve the inverse problem, i.e. which mass gives observed stellar motions and bridges the gap between SMBH and stellar-mass black holes.  However, the Extremely Large Telescope is coming with extremely large challenges. 

Existing PSF fitting algorithms face strong PSF variations across FoV in SCAO mode (which will be used for the first light of MICADO, before the launch of MORFEO - MCAO mode). This challenge is usually avoided, e.g. \citet{2021ASPC..528..167S}, \citet{10.1093/mnras/staa869} used the assumption that IMBH lives in the center of the stellar cluster, simulated the central small part 5''x5'', considered PSF as a constant, then used {\tt\string Starfinder} and {\tt\string DAOPHOT/ALLSTAR} respectively for PSF fitting. However, realistic N-body simulation {\tt\string Dragon-II} \citep{2023arXiv230704805A} shows that one cannot strictly imply the central position of an IMBH inside the stellar cluster, as IMBH can move around the center of mass. This fact highlights the importance of taking PSF variations into account because the IMBH can have a projection offset from the AO reference star which is hard to predict.
\section{Quality metrics}
In our work, 
interpolated PSF quality metrics are shown from four points of view, which are described in the following subsections. 
\subsection{PSF as image}
For the calculation of image-comparison metrics, we use {\tt\string scikit-image} \citep{scikit-image} Python package. We consider mean
absolute percentage error (MAPE),
root mean squared error (RMSE),
normalized RMSE (NRMSE),
peak signal-to-noise ratio (PSNR).
\subsection{PSF as multidimensional Probability Density Function}
The signal from the investigated point source is a wave of electromagnetic (EM) field with the amplitude and the phase. Following the uncertainty principle for energy $\Delta E \Delta t \geq \hbar$ and $\Delta E = \hbar \omega \Delta N$ we obtain the uncertainty principle for phase $\Delta N \Delta \phi \geq   1$, which shows that we can not measure the phase of the EM field for a certain number of photons. Existing detectors of incoming energy measure voltage or current caused by a certain number of absorbed photons. When EM waves propagate through atmospheric turbulence and aberrating optical elements, the wavefront changes and a number of counted photons changes as well, because the constructive and destructive interference in practice constrains whether the photon will hit the CCD or not. PSF is by design a conditional Probability Density Function (PDF) of a photon hitting a pixel with the condition being that it has reached the CCD. 
In this definition, PSF has the following properties:
\begin{enumerate}
\item Normalized: \begin{equation} \int^{\infty}_{-\infty} PSF(x,y)=1 \end{equation}
\item Positive: \begin{equation} PSF \geq 0 \end{equation} 
\item Not transformable back to wavefront domain. 
\end{enumerate}
For a certain instrument, PSF depends on two groups of parameters: more or less static (e.g. central wavelength of the passband, distance from reference star) and dynamic parameters (e.g. atmosphere model, the temperature of optical elements, weather conditions, zenith angle), which depend on time.
In the probabilistic view, it means that PSF is a multidimensional PDF in the first approximation and a multidimensional stochastic (random) process in the second one
(e.g. as multidimensional Gaussian Processes).
For the comparison of two PDFs, divergence-family metrics (separation of probability distributions in the statistical manifold): Kullback-Leibler Divergence (KL div), squared Hellinger distance, Jensen-Shannon divergence, and chi-squared divergence, can be used. KL div keeps the positivity feature, but does not keep symmetry in the sense $f(A,B)=f(B,A)$ and consequently can not be treated as a distance (metric) and does not form a metric space, i.e. in some sense does not provide interpretable comparison of the PSFs.
Additionally, PSFs can be fitted by PDFs (e.g. 2D or 3D Gaussian, Moffat), and parameters of distribution can be compared. The quantiles of PSF are metrics as well, and partially similar to Encircled Energy (EE).
\subsection{Astronomical metrics}
The python package POPPY \citet{2016ascl.soft02018P} provides the implementation of the wide-used range of optical metrics: strehl ratio, EE, full~width~at~half~maximum \\ (FWHM), centroid of PSF.
\subsection{Distinguishing of 2 close stars}
In the perspective of choosing an interpolation method for high-precision astrometry, we use the separation of the two closest stars as the specific metric. In the particular case of crowded fields (e.g. stellar clusters), one of the most significant challenges is to separate overlapping PSFs. The search of IMBH is based on proper motion measurements for the stars relatively close to the dark central mass. Since \citet{2023MNRAS.526..429A} modern simulations manifest the birth of IMBH from massive stars and stellar-mass black holes, we expect to see wide wings of massive bright stars in a mixture with faint stars. For the separation of the two closest stars, we propose to use Expectation-maximization (EM) algorithm \citet{477e7e2b-4ded-3369-981e-9b40850a2701}. This algorithm appeals to work for data with hidden parameters, so-called latent variables. In our case, e.g. wavefront is the latent variable, since we work with amplitudes of Fourier-transformed aperture function.



\section{Conclusion}
We showed the scope of PSF quality metrics for PSF interpolation, ePSF fitting, and comparison of PSFs across FoV. For stellar-crowded photometry, we proposed to use distinguishing of 2 closest stars as an additional metric.






\acknowledgements I would like to appreciate the following people for discussion of and advice on my current work: Igor Chilingarian, Anton Afanasiev, Konstantin Malanchev, Nadine Neumayer, Kieran Leschinski, Mikhail Hushchyn, Roy van Boekel, Markus Feldt, Ren\'{e} Andrae, Kirill Voronin, Kirill Shiianov, Ivelina Momcheva, Morgan Fouesneau, Georgii Strukov. 
\bibliography{P902}  

\end{document}